\documentclass{elsart}

\usepackage{epsfig}

\usepackage{amssymb}
\usepackage{subfigure}

\begin{document}

\begin{frontmatter}

\title{A high efficiency, low background detector for measuring pair-decay branches in nuclear decay}

\author[nsclmsu,jina,msupa]{C. Tur}
\author[wmu,jina]{A. H. Wuosmaa}
\author[nsclmsu,jina]{S. M. Austin}
\author[nsclmsu,msupa]{K. Starosta}
\author[nsclmsu]{J. Yurkon}
\author[nsclmsu,jina,msupa]{A. Estrade}
\author[wmu]{N. Goodman}
\author[wmu]{J. C. Lighthall}
\author[nsclmsu,jina,msupa]{G. Lorusso}
\author[wmu]{S. T. Marley}
\author[wmu]{J. Snyder}
\corauth[cor1]{Corresponding Author: tur@nscl.msu.edu,
(517)333-6312, fax (517)324-8135}
\address[nsclmsu]{\scriptsize{National Superconducting Cyclotron Laboratory, Michigan State University, East Lansing, MI 48824-1321, USA}}
\address[wmu]{\scriptsize{Department of Physics, Western Michigan University, Kalamazoo, MI 49008-5252, USA}}
\address[jina]{\scriptsize{Joint Institute for Nuclear Astrophysics, Michigan State University, East Lansing, MI 48824, USA}}
\address[msupa]{\scriptsize{Department of Physics and Astronomy, Michigan State University, East Lansing, MI 48824, USA}}

\begin{abstract}
We describe a high efficiency detector for measuring electron-positron 
pair transitions in nuclei. The device was built to be insensitive to 
gamma rays and to accommodate high overall event rates.  The design was 
optimized for total pair kinetic energies up to about 7 MeV.
\end{abstract}

\end{frontmatter}

\section{Introduction}
During stellar helium burning, the ``triple alpha'' ($3\alpha$) reaction
involving the excited $0^+$ state at
7.65 MeV (the so-called ``Hoyle state'') in $^{12}$C converts
alpha particles to $^{12}$C. Certain stellar processes are sensitive
to the  $3\alpha$ reaction;  current models of stellar evolution
require that its rate must be known to within about
5\% to describe these processes accurately \cite{aus05}. The reaction 
rate is inversely proportional to the pair branch ($\Gamma_\pi/\Gamma$), 
of the Hoyle state, the fraction of the time that, once formed, this 
state decays to the ground state by the emission of an electron-positron 
pair. This branch is presently known with an accuracy of 9.2\% 
~\cite{alb60,alb77,rob77} and is by far the largest 
uncertainty in determining the rate.

\begin{figure}
\centering
\includegraphics[width=0.95\textwidth]{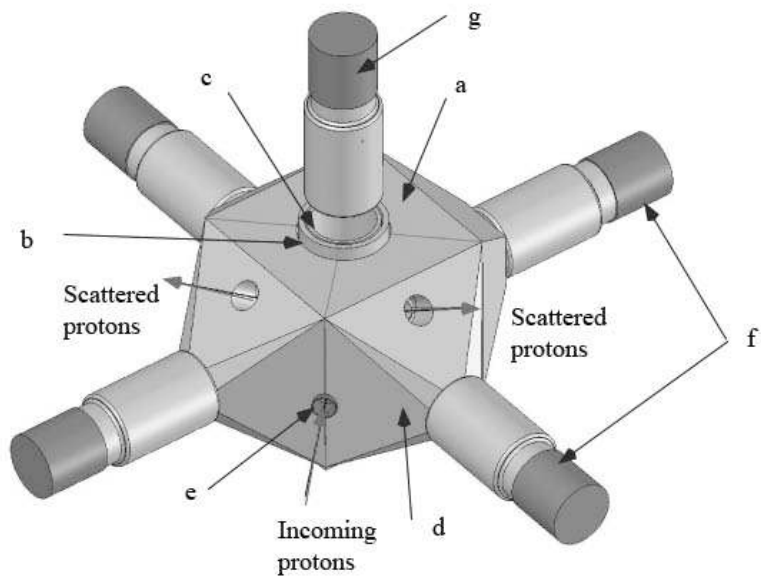}
\caption{Schematic view of the detector.  The letters correspond
to those in Table \ref{detectorDim}. The Si detectors are not shown.}
\label{detector}
\end{figure}

\begin{figure}
\centering
\includegraphics[width=0.8\textwidth]{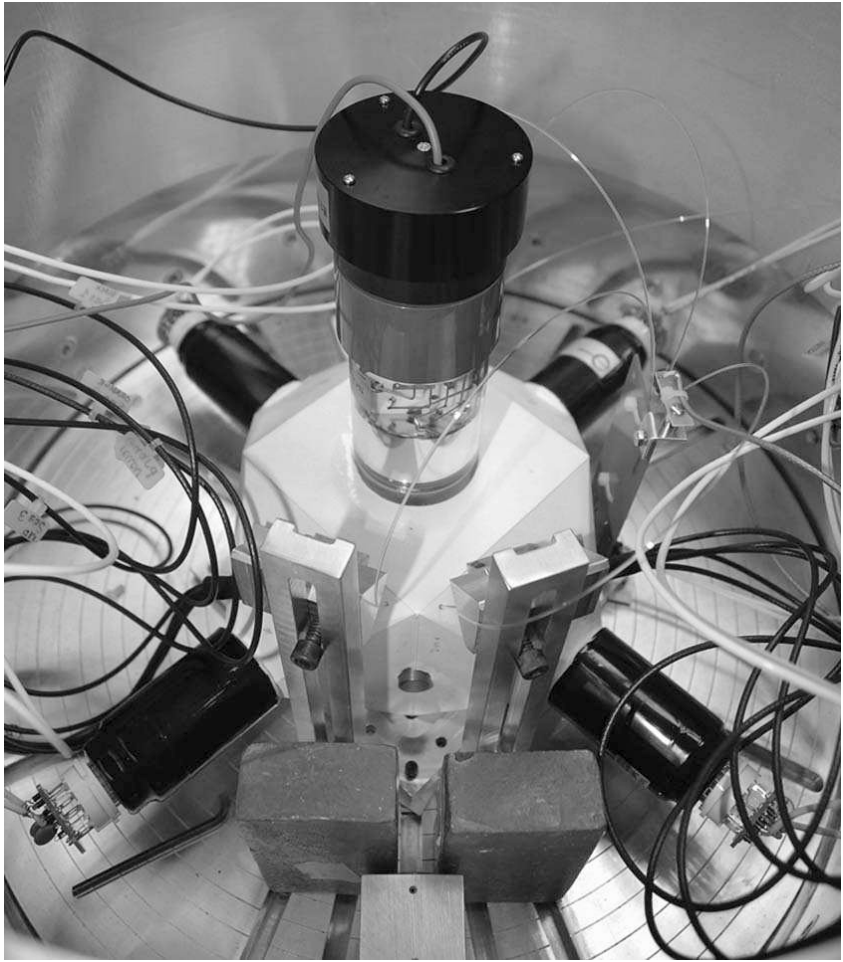}
\caption{Photograph of the detector installed in the 
scattering chamber. Lead shields were placed in front of the beam entrance hole
to reduce background due to the scattering of the beam on the collimators.}
\label{detector2}
\end{figure}

The detector described here was designed to measure
$\Gamma_\pi/\Gamma$ to within about 5\%, resulting in a determination
of the $3\alpha$ rate to about 6\%.  The experiment is simple in
principle: the 0$^+_2$ excited state in $^{12}$C is excited by inelastic 
proton scattering at a bombarding energy of 10.4 MeV.
 The population of the state is tagged by the observation of scattered
protons detected using two Si PIN diodes placed at $125^{\circ}$ in the 
laboratory with respect to the beam direction. The choice of the scattering 
angle takes advantage of a resonance in the excitation function of
the $^{12}$C(p,p$^{'}$)$^{12}$C(7.65 MeV) reaction to maximize the count
rate \cite{swi66}.  Positron-electron pairs from the decay
of the 0$^+_2$ state are detected in coincidence with protons using an array of
plastic scintillators.  The ratio of the proton-pair coincidence rate
to the proton singles rate then gives the branching ratio, after correction 
for detection efficiency, random coincidences and gamma-ray
contamination. The principal difficulty of this measurement is that
$\Gamma_\pi/\Gamma$ is extremely small, about $6.7\times 10^{-6}$, and
hence sensitive to even small backgrounds for the e$^+$-e$^-$ pairs. The
dominant physical background is the much more probable cascade 
gamma decay of the Hoyle state which possesses a branching ratio of
$\Gamma_\gamma / \Gamma = 4.1\times 10^{-4}$.  Excitation of
the $2^{+}$ state at 4.44 MeV in $^{12}$C and of states in target 
impurities, e.g. $^{28}$Si, are other sources of background. 
Additional (random) backgrounds can be produced by interactions of the 
proton beam with collimators, elements of the detector, and the beam stop. 
Further details on the experiment are given in References ~\cite{aus05} 
and ~\cite{tur06}.

Section II describes the detector constructed for this purpose
with special attention to features important for the reduction of
backgrounds. In sections III and IV the results of GEANT4 simulations are 
discussed and compared with the observed performance of the detector.

\section{Detector properties and construction}

The conceptual design of the detector is shown in Figure
~\ref{detector}.  Figure ~\ref{detector2} shows a photograph of the 
detector installed inside the scattering chamber.  The
active elements of the detector, consisting of a thin tube surrounded by a 
square block segmented into four quadrants, were constructed of Bicron BC404
plastic scintillator.  The energies deposited by the electrons and positrons
in the quadrants of the scintillator block were read out using photomultiplier 
tubes (PMTs) attached to each quadrant with a light guide constructed 
of BC800 plastic. The scintillator tube was coupled to a PMT 
using a flexible silicone pad. By requiring coincident events in
one or more quadrants and the central tube, gamma-ray backgrounds
were suppressed by the low sensitivity of the 
thin scintillator tube to gamma rays. In order to shield the active
elements of the scintillator array from protons scattered from the
target, a 2 mm thick tube of inert plastic (acrylic) surrounded
the target inside of the active tube as shown in
Fig.~\ref{detector}. Holes were bored in this central tube to allow
protons in the beam, or protons that were to be detected in the
silicon detectors to emerge. Table ~\ref{detectorDim} summarizes
the materials used in the construction of the detector.

\begin{table*}
\centering
\caption{Detector materials and dimensions--letters refer 
to Figure ~\ref{detector}. All circular dimensions are diameters, with 
OD to mean ``outer diameter'' and ID, ``inner diameter''.}
\label{detectorDim}
\begin{tabular*}{0.95\textwidth}{ccc}
\hline 
\hline 
Object & Material & Dimensions\\
\hline 
Scintillator block (a) & Bicron BC404 & 10$\times$10$\times10$ cm$^3$,\\
   && central hole 50 mm ,\\
   && four quadrants\\
Scintillator tube (b) & Bicron BC404 & height 105 mm,\\
   && 40 mm ID, 46 mm OD,\\
   && all surfaces polished\\
Absorber tube (c) & Clear acrylic & height 105 mm,\\
   && 34 mm ID, 38 mm OD,\\
   && all surfaces polished\\
Light guides (d) &Bicron-800\\
Beam hole liner (e) & Tantalum & 11 mm ID, 12 mm OD\\
Paint & F113 Epoxy with TiO2 & about 0.25 mm thick\\
Quadrant PMTs (f) &   Electron tubes 9845B&3.2 cm photocathode\\
Tube PMT (g) &Hamamatsu R6231&4.6 cm photocathode\\
Si detectors & Si PIN diodes & 1 mm thick, 3 sub-sectors\\
\hline
\end{tabular*}
\end{table*}

The absorber and scintillator tubes, as well as the surrounding 
scintillator blocks, were mounted on an aluminum 
base plate. The surfaces of the scintillator quadrants 
were covered with white epoxy paint so as to provide light insulation 
between the quadrants, as well as to improve light collection.
The segmentation of the square block into quadrants improves the rate 
capability of the device by limiting pulse pile up.

The beam entrance and exit holes were lined with tantalum tubes which
prevent beam particles from entering the scintillators. The beam holes 
were also painted with white epoxy paint before insertion of the liners 
so as to prevent the absorption of scintillation light by the dark liner.

The silicon detectors were 1 mm thick PIN diodes segmented into three sectors 
and are described in Reference
~\cite{eve93}. The solid angle for protons is defined by two 4 mm
diameter holes in the acrylic tube at $125^\circ$ in the
laboratory with respect to the beam direction. These holes extend
through the scintillator tube with a diameter of 12.7 mm; the holes 
through the adjacent quadrants have a diameter of 12.7 mm, widening to 16 mm 
at the light guides to prevent the exiting particles from scattering 
back into the scintillators. The size of the apertures was chosen to keep the
kinematic spread in proton energy comparable to the intrinsic Si
detector resolution.  The acceptance solid angle can be further
restricted by tantalum apertures placed in front of the Si detectors,
so as to reduce the kinematic spread.

\section{GEANT4 simulations}

We have carried out Monte Carlo simulations of the scintillator
detectors using the code GEANT4~\cite{gea07}. The aim of the simulations is 
to guide the design, as well as to understand the response of the
scintillators to electrons, positrons, and gamma rays.  This
understanding is crucial to the determination of important operational
characteristics such as pair efficiency, gamma-ray efficiency, and
suppression.  In practice, these properties are determined from
comparisons between data and simulation for isolated, well-identified
transitions. For the gamma-ray cascade of the $^{12}$C(0$^+_2$) state,
the response to the 3.21 MeV gamma ray is determined from the Monte Carlo 
simulation.

\subsection{Elements of the GEANT4 simulation}

The GEANT4 simulation includes all the active elements of
the plastic scintillator detectors.  These elements are modeled
with a realistic detector geometry.
The central, inactive tube is also included in the simulation
as particles and gamma rays deposit energy in this element as well.
The simulation provides the energy deposited in each volume included in the
experimental geometry.  Elements not included are the light guides,
aspects of light collection and light attenuation, and the response of the 
PMTs. Experimental resolution effects are treated later in the
simulation process as described below.  

Electron-positron pairs from E0 transitions were simulated using 
the well-known energy and angular correlations given by the Born
approximation~\cite{opp41}:

\begin{equation}
\frac{dN}{d\epsilon}=(\epsilon^{2}-1)^{1/2}\times(\epsilon'^{2}-1)^{1/2}\times
(\epsilon\epsilon'-1)
\label{energyCorr}
\end{equation}

where $\epsilon$ and $\epsilon'$ are the total energies of the electron and
positron. The angular correlation between the electron and positron is
given by:

\begin{equation}
\frac{d^2N}{d\epsilon dcos\theta} = \frac{dN}{d\epsilon}
\times(1 + \alpha\times cos{\theta})
\label{angularCorr}
\end{equation}

where $\alpha$ is the E0 anisotropy factor derived using the Born 
approximation (see Appendix C of Reference ~\cite{hof90}), and is
given by:

\begin{equation}
\alpha = \frac{(\epsilon^{2}-1)^{1/2}\times(\epsilon'^{2}-1)^{1/2}}{(\epsilon\epsilon'-1)}
\label{anisotropy}
\end{equation}

The response of the detector to pairs from the decay of the
$^{12}$C($0^+_2$; 7.65 MeV) and $^{16}$O(0$^+_2$; 6.05 MeV) states
was studied.  The 0$^+_2$ state in $^{16}$O provides a
convenient calibration signal as it has a 100\% e$^+$-e$^-$ decay
branch to the $^{16}$O ground state.  All particles were followed and
their energy losses monitored until they either stopped, annihilated, or
left the detector.

\begin{figure}
\centering
\includegraphics[width=0.6\textwidth]{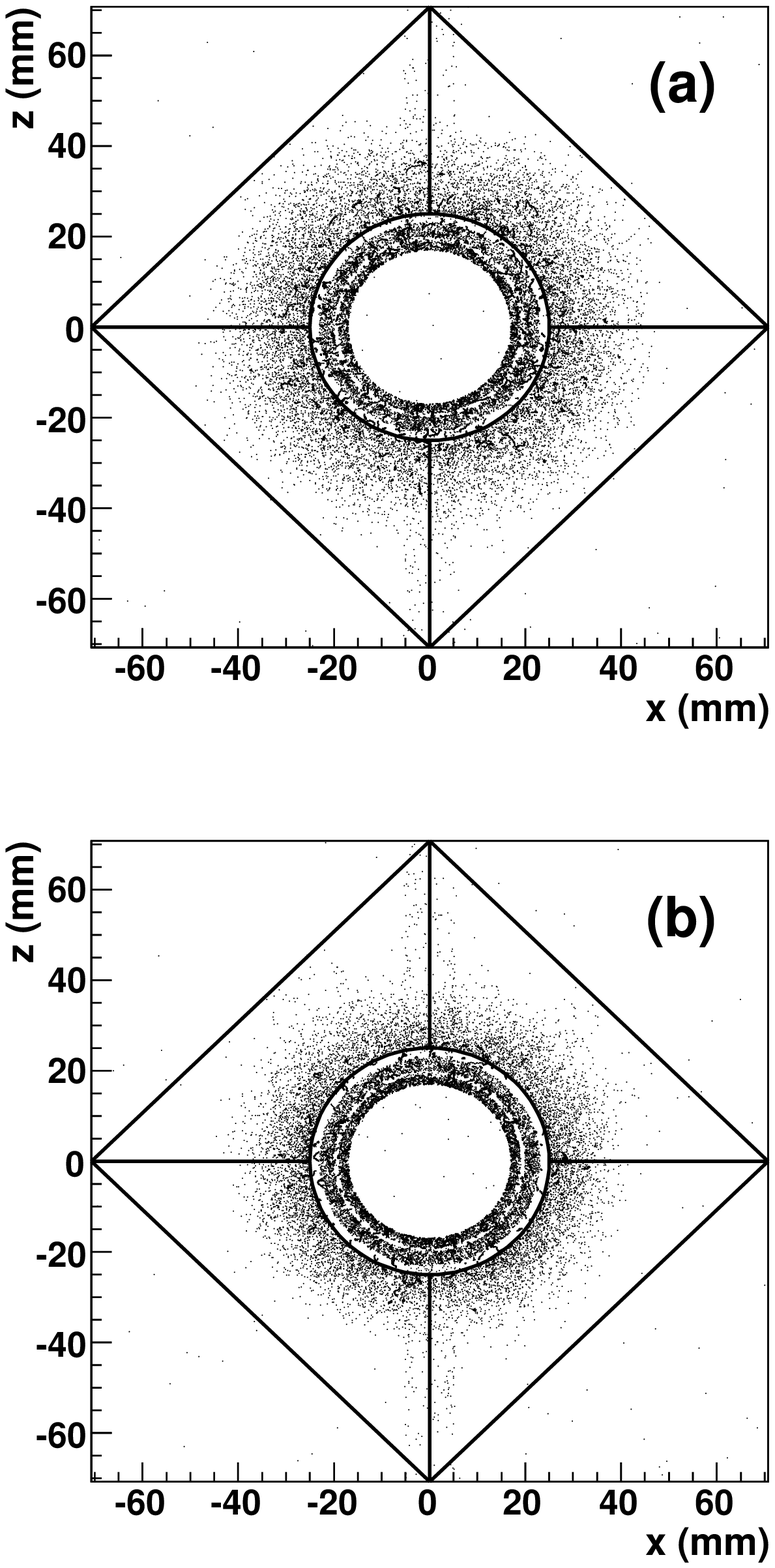}
\caption{The x and z coordinates (in mm) of the points 
where the electrons and positrons from the pair decay of the (a)
$^{12}$C(0$^+_2$; 7.65 MeV) and (b) $^{16}$O(0$^+_2$; 6.05 MeV) states
stop for $10^{4}$ generated pairs. The thick lines show the contours 
of the four scintillator blocks surrounding the acrylic tube and the 
scintillator tube. For those particles 
which step outside the detector, that first step outside is shown. The 
positrons which annihilate in flight are not represented in the plots.}
\label{stoppingRange}
\end{figure}

\begin{figure}
\centering
\includegraphics[width=1.0\textwidth]{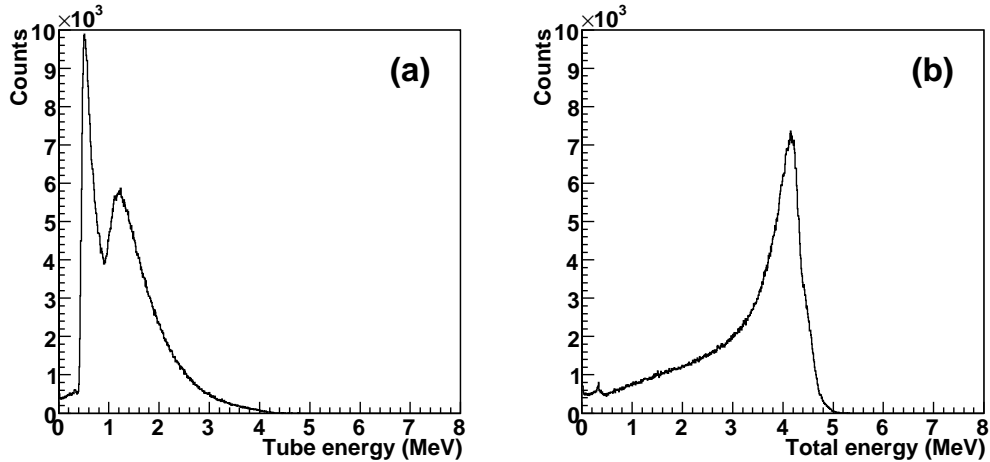}
\caption{(a) Simulated spectrum of energy deposited in the active
scintillator tube by 10$^6$ e$^+$-e$^-$ pairs from the decay of the
$^{16}$O(0$^+_2$; 6.05 MeV) state. (b) Spectrum of total energy
deposited in all active detector elements by pairs produced from the
decay of the $^{16}$O(0$^+_2$) state.}
\label{Edep}
\end{figure}

\begin{figure}
\centering
\includegraphics[width=1.0\textwidth]{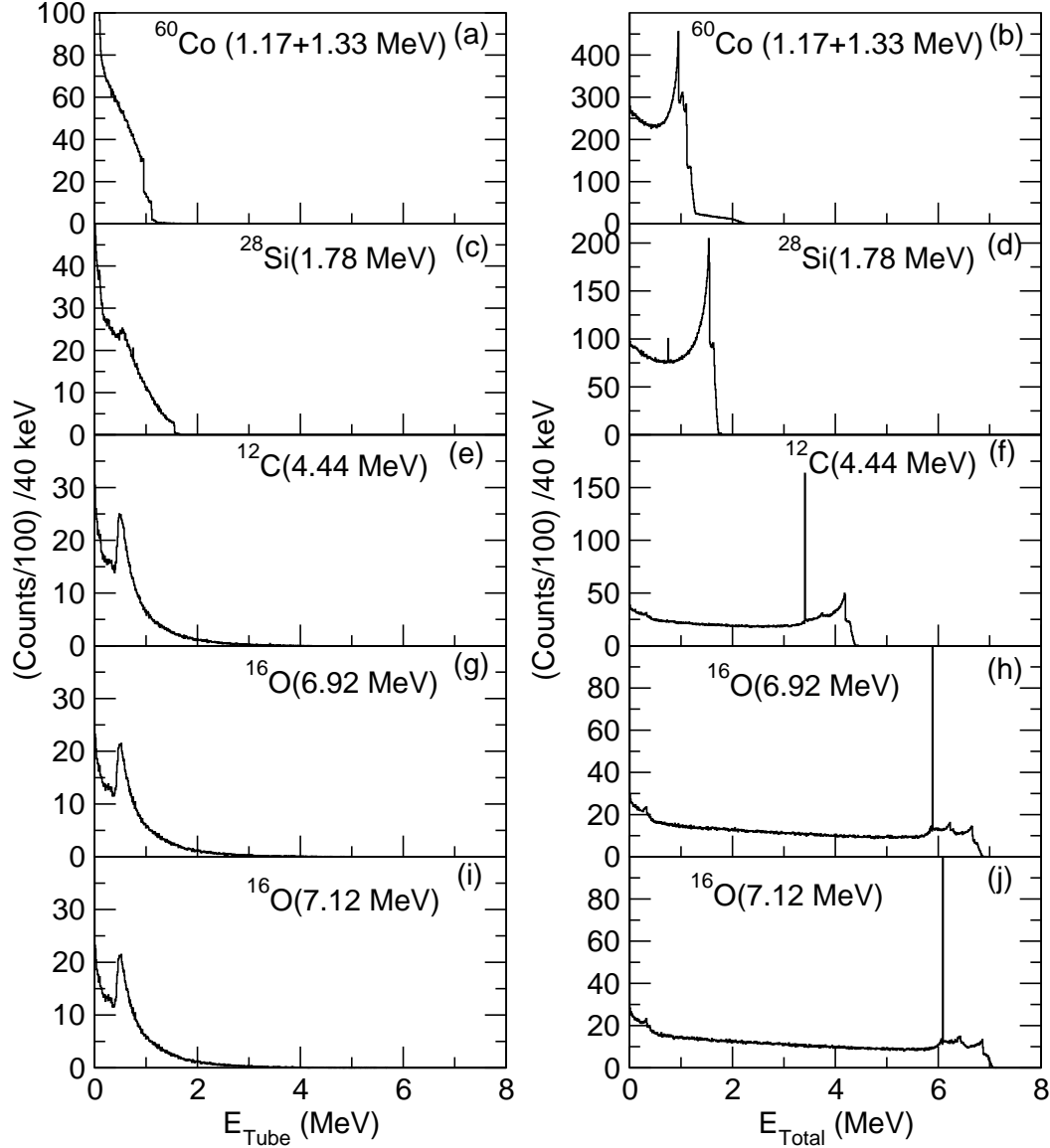}
\caption{Simulated energy-deposition  spectra for $10^{7}$ generated events 
for the tube (left hand side) 
and all active detector elements (right hand side) for gamma rays 
with energy (a),(b):1.17+1.33 MeV ($^{60}$Co); (c),(d): 1.78 MeV ($^{28}$Si); 
(e),(f): 4.44 MeV ($^{12}$C); (g),(h): 6.92 MeV ($^{16}$O) and 
(i),(j): 7.12 MeV ($^{16}$O). The spike at 1.022 MeV below the total
energy corresponds to escape of both 511 keV annihilation gamma rays.}
\label{gammaSims}
\end{figure}

The response of the detector to 4.44 and 3.21 MeV gamma rays (the
energies produced in the gamma-ray cascade decay of the
$^{12}$C($0^+_2$) level), gamma rays from radioactive $^{60}$Co and
$^{137}$Cs sources, and a variety of other transitions in $^{28}$Si
and $^{16}$O were simulated.  In all cases, the photons were emitted
from the target center and distributed isotropically throughout the entire
solid angle.  In the case of two-gamma-ray cascades, the photons were
each emitted isotropically and the angular correlation between them
was ignored. For positrons, electrons, and gamma rays the low-energy
processes of GEANT4 were used to extend the validity range of the
coded electro-magnetic interactions and cover the full range of
experimental energies (see Chapter 12 of Reference ~\cite{gea07}). These
processes include annihilation in flight for positrons.

\subsection{e$^+$-e$^-$ response and efficiency}

\begin{figure}
\centering
\includegraphics[width=0.8\textwidth]{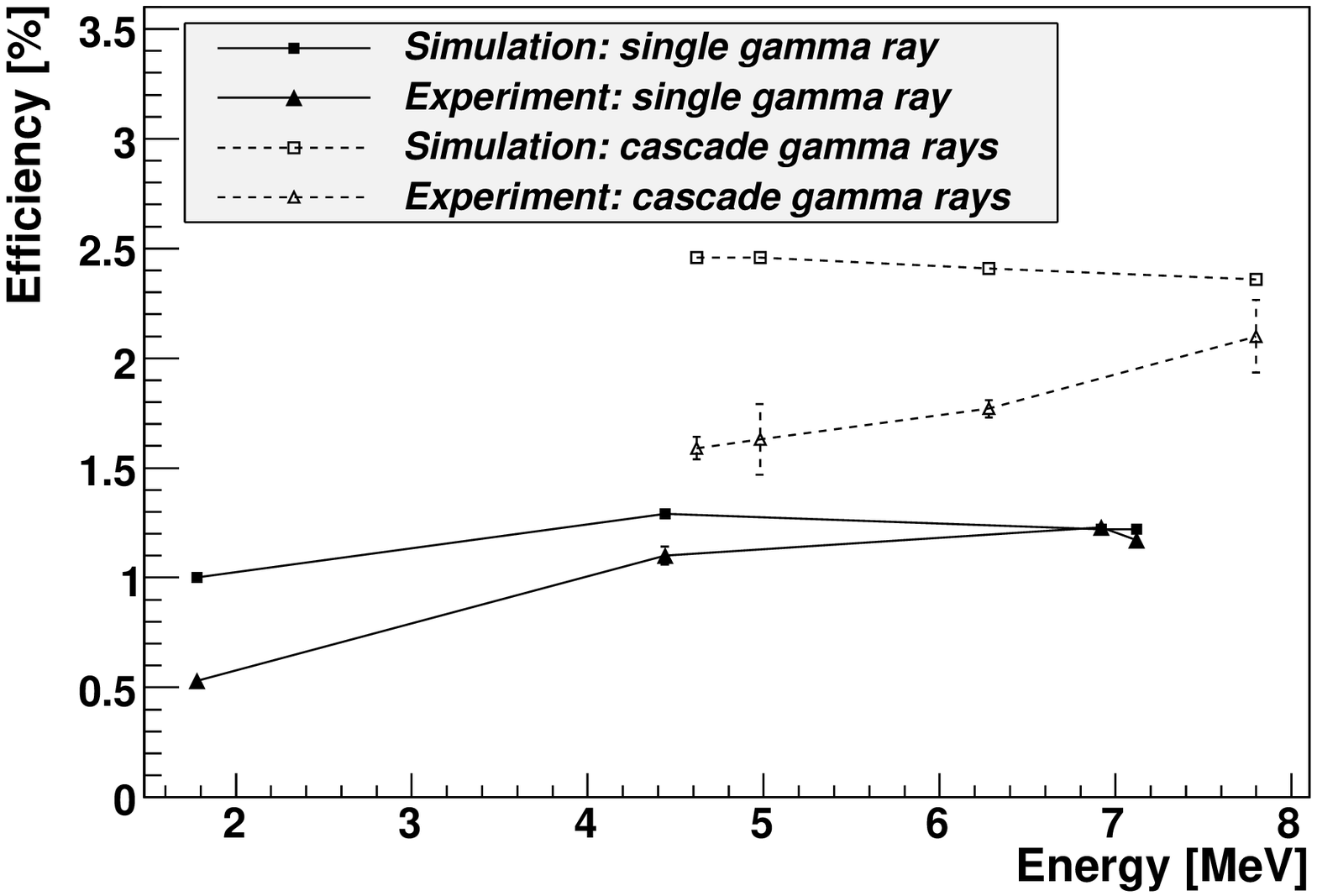}
\caption{Simulated and experimental detection efficiencies for single 
gamma-ray ($^{28}$Si(1.78 MeV), $^{12}$C(4.44 MeV), 
$^{16}$O(6.92 MeV), and $^{16}$O(7.12 MeV)) and cascade gamma-ray transitions 
($^{28}$Si(2.84 + 1.78 MeV), $^{28}$Si(3.20 + 1.78 MeV), 
$^{28}$Si(4.50 + 1.78 MeV), and $^{28}$Si(6.02 + 1.78 MeV)). 
See the text for an explanation of the discrepancies between the simulated 
and the corresponding experimental efficiencies. For cascade
gamma-ray transitions, the sum of the energies of the two gamma rays is 
shown on the x-axis. The error bars are purely statistical. Where not visible,
the uncertainties are smaller than the size of the symbols. The 
statistical uncertainties for the simulations are negligible.}
\label{gammaEff}
\end{figure}

\begin{figure}
\centering
\includegraphics[width=0.8\textwidth]{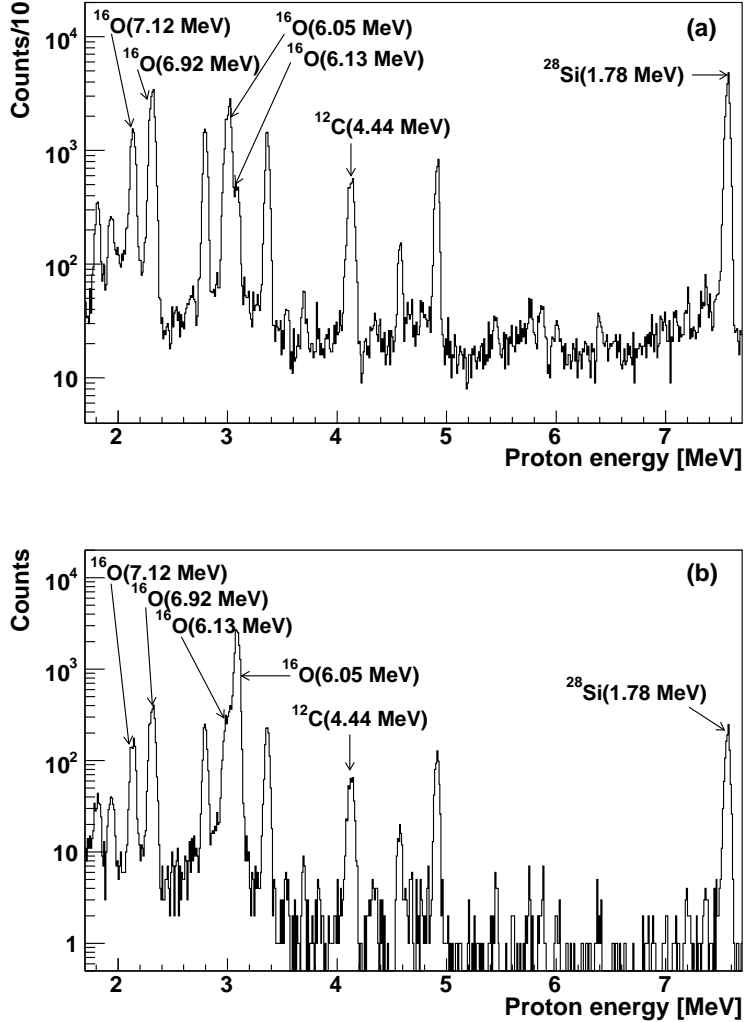}
\caption{Proton energy for one segment of one of the silicon detectors 
showing (a) the singles and (b) the coincidence spectra in the energy range 
1.7-7.7 MeV for the SiO$_{2}$ target. The single gamma ray transitions for 
which the estimated experimental efficiencies are shown on 
Figure ~\ref{gammaEff} are labeled in addition to the $^{16}$O(6.05 MeV)
and the $^{16}$O(6.13 MeV) transitions. All peaks have been identified; the
majority are from Si. The statistics in (a) reflect a down-scale 
factor of 10.}
\label{gammaSi}
\end{figure}

\begin{figure*}
\centerline{
\mbox{\includegraphics[width=0.497\textwidth]{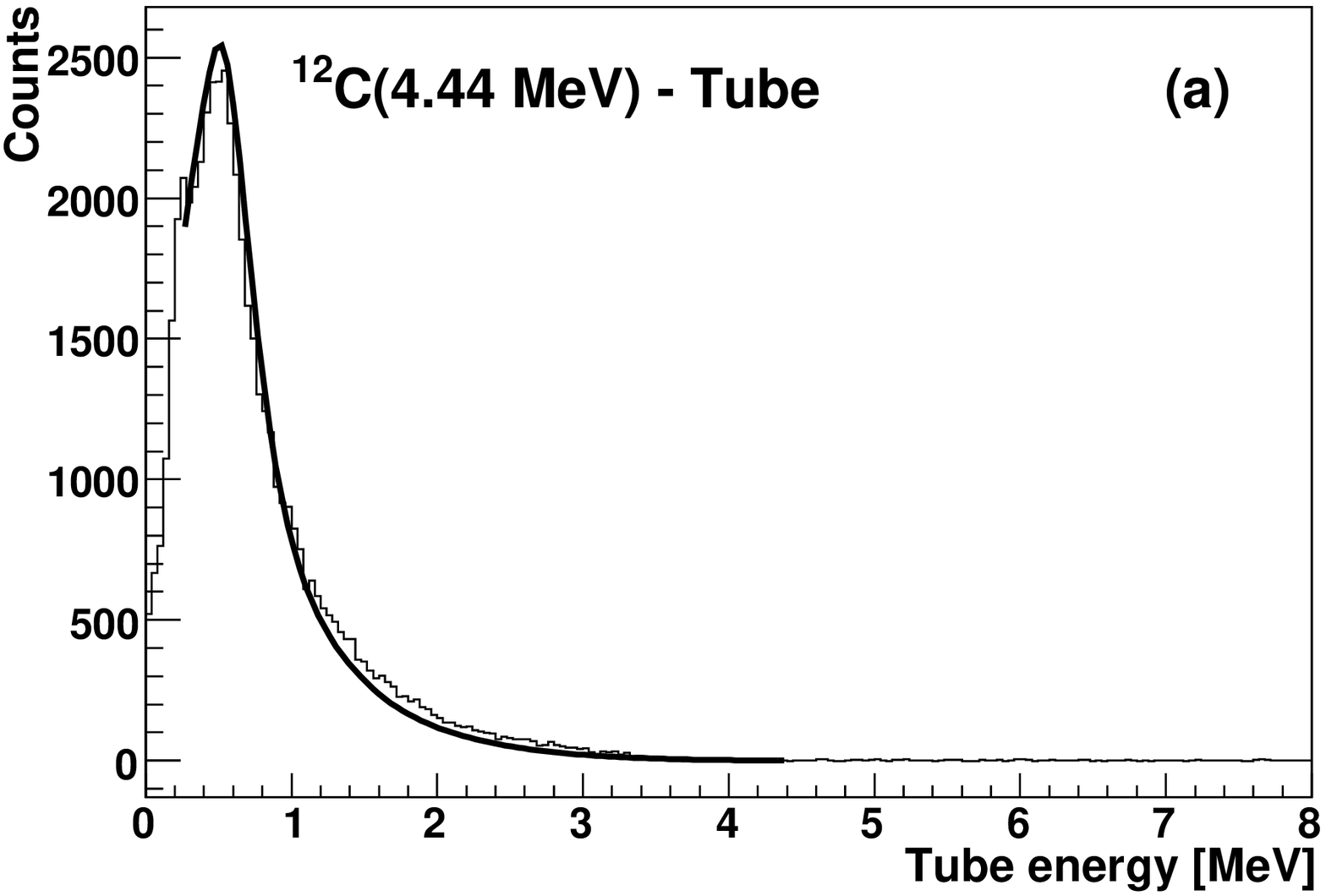}}
\mbox{\includegraphics[width=0.497\textwidth]{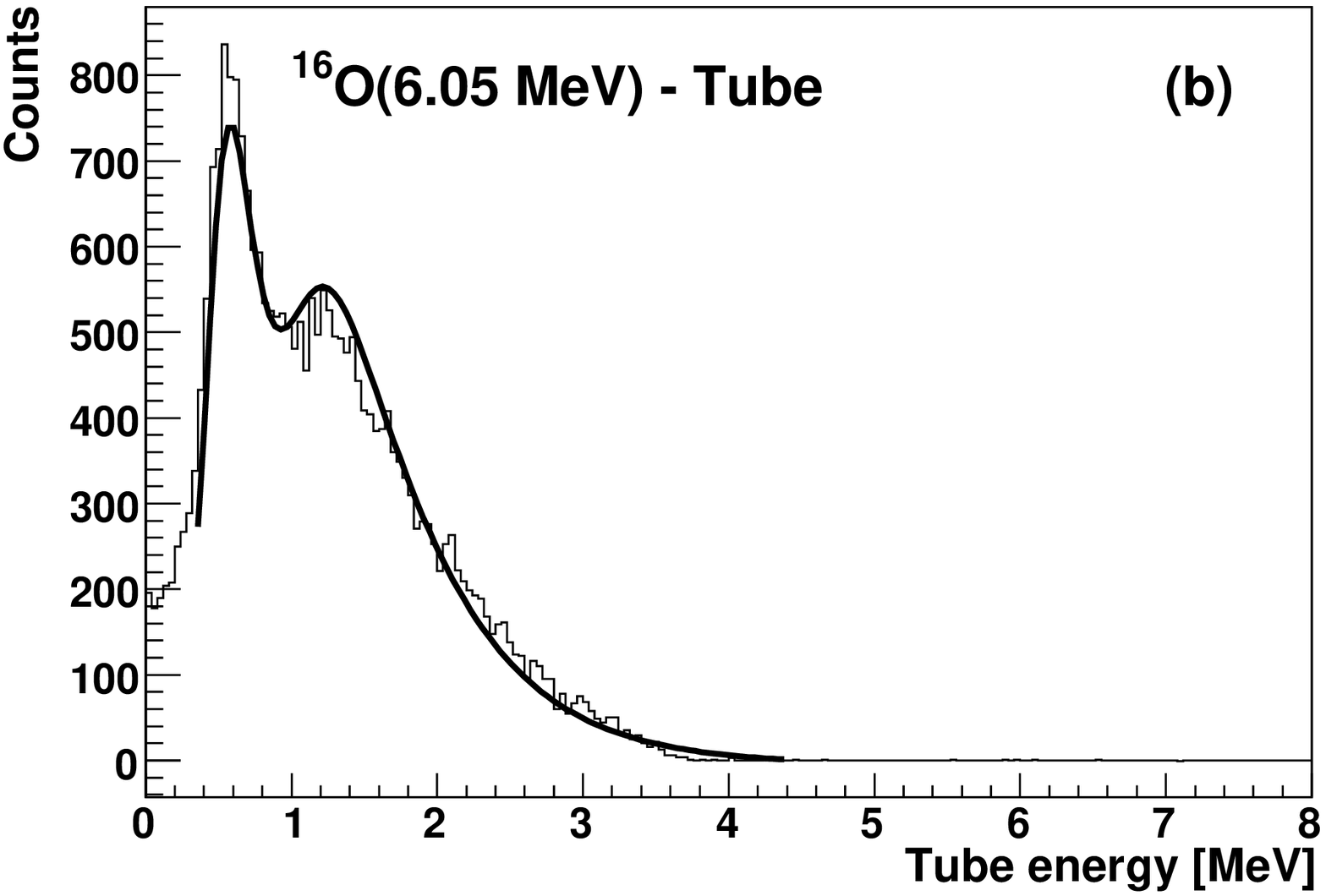}}
}

\caption{Two-point tube calibration using (a) the peak at 0.5 MeV for $^{12}$C(4.44 MeV) and (b) the one at 1.2 MeV for $^{16}$O(6.05 MeV). The thick dark 
lines show a fit to the data using an empirically determined detector 
resolution function for the tube (see text).}
\label{calibration}
\end{figure*}

\begin{figure}
\centering
\centerline{
\mbox{\includegraphics[width=0.497\textwidth]{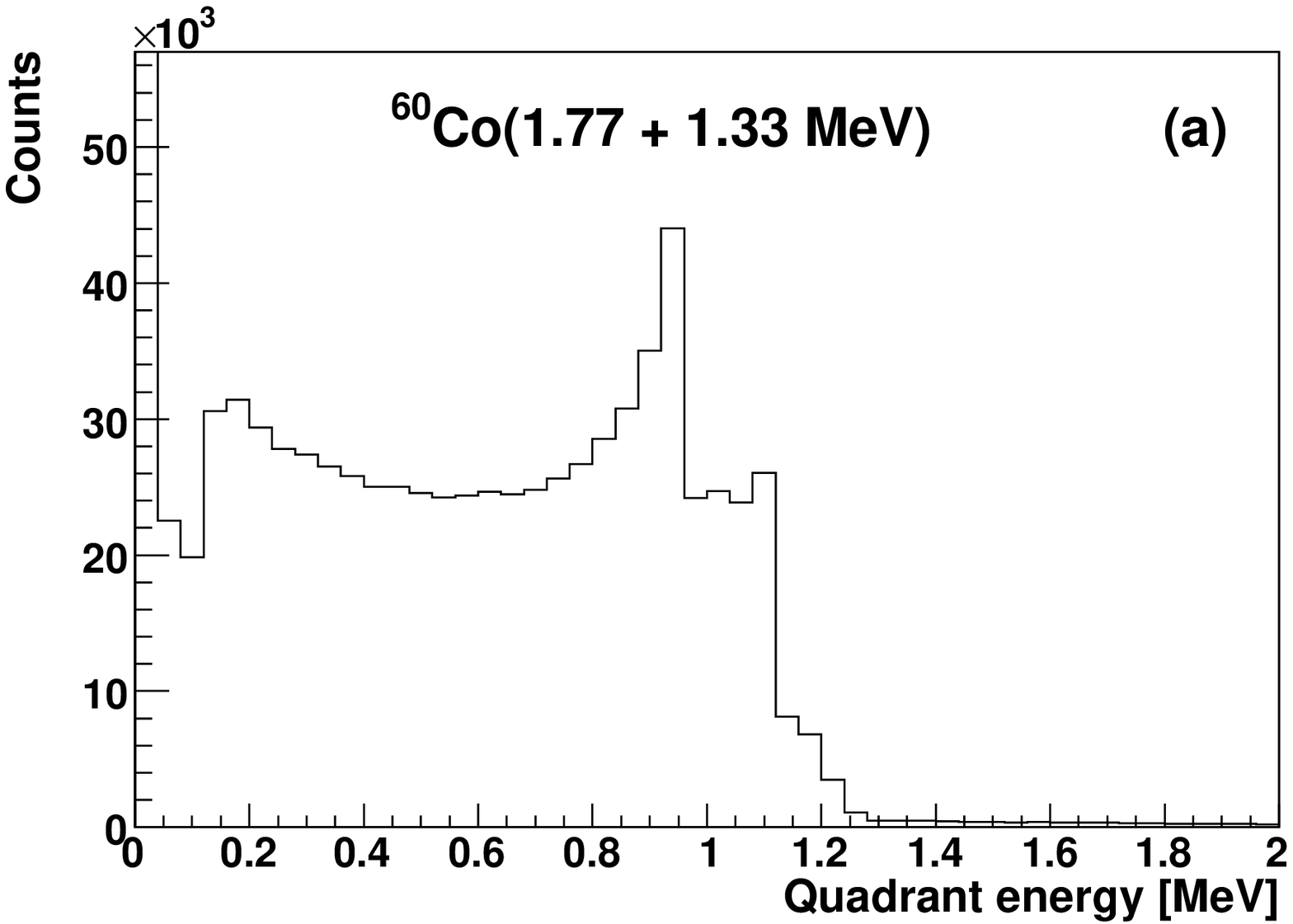}}
\mbox{\includegraphics[width=0.497\textwidth]{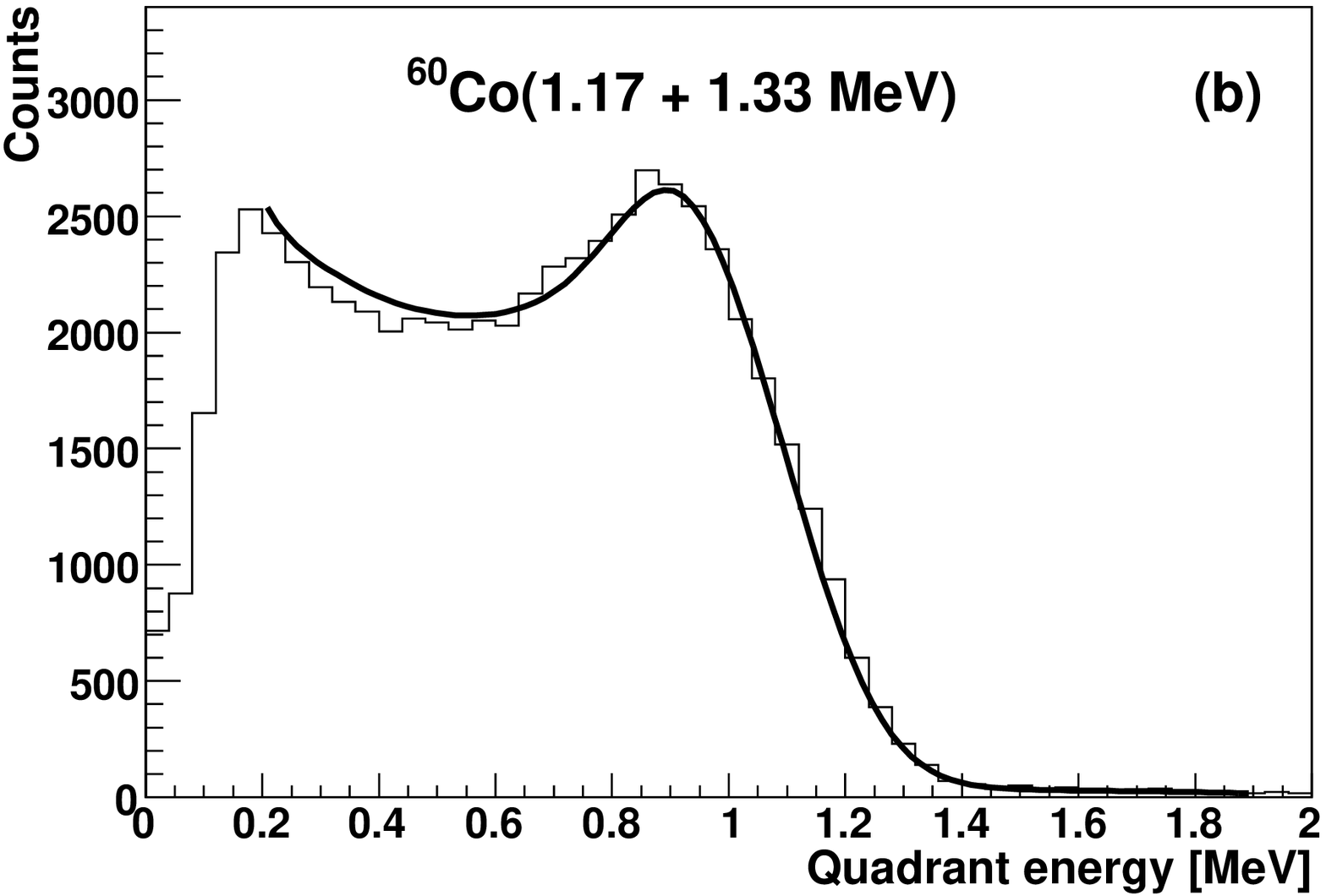}}
}
\centerline{
\mbox{\includegraphics[width=0.497\textwidth]{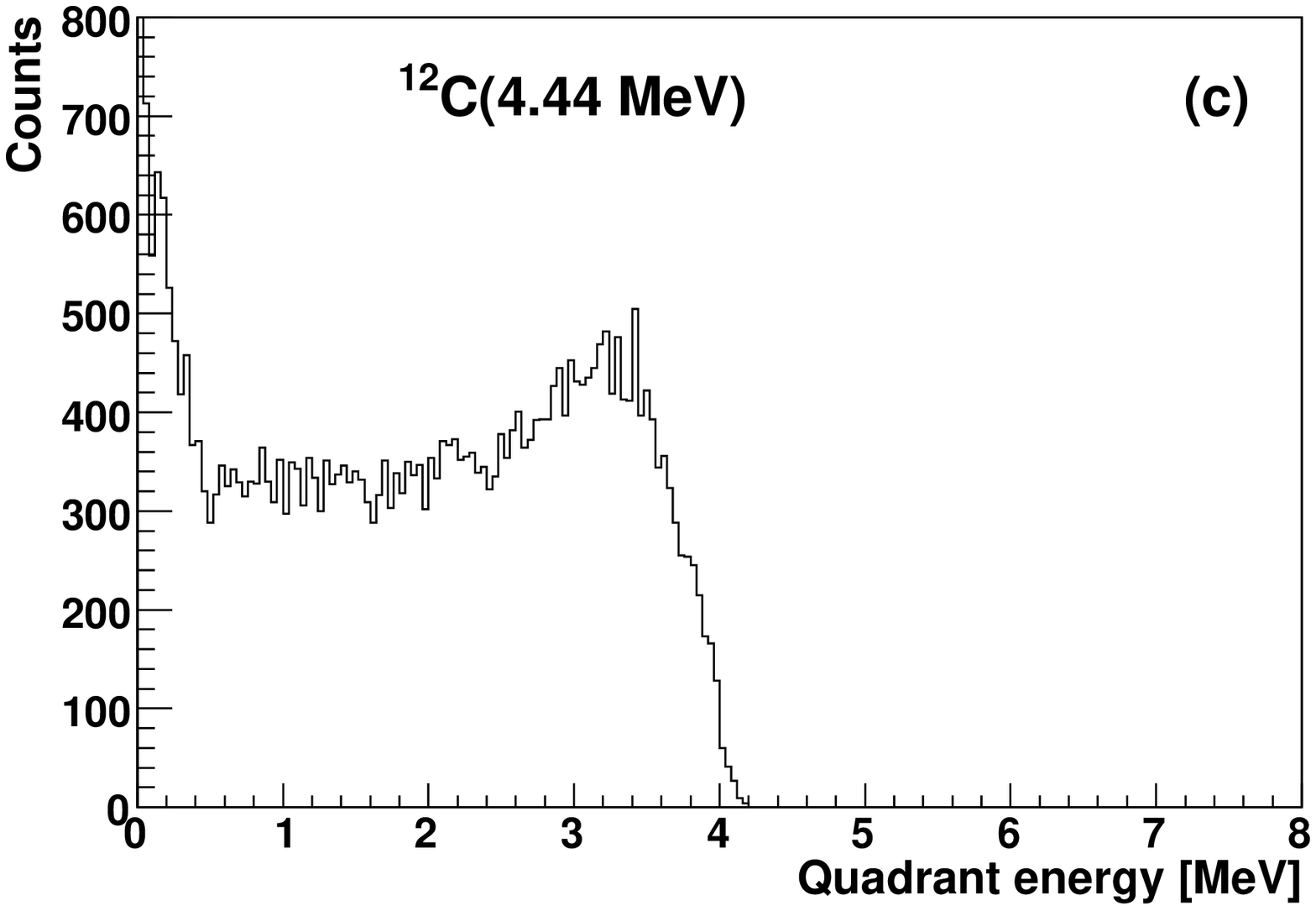}}
\mbox{\includegraphics[width=0.497\textwidth]{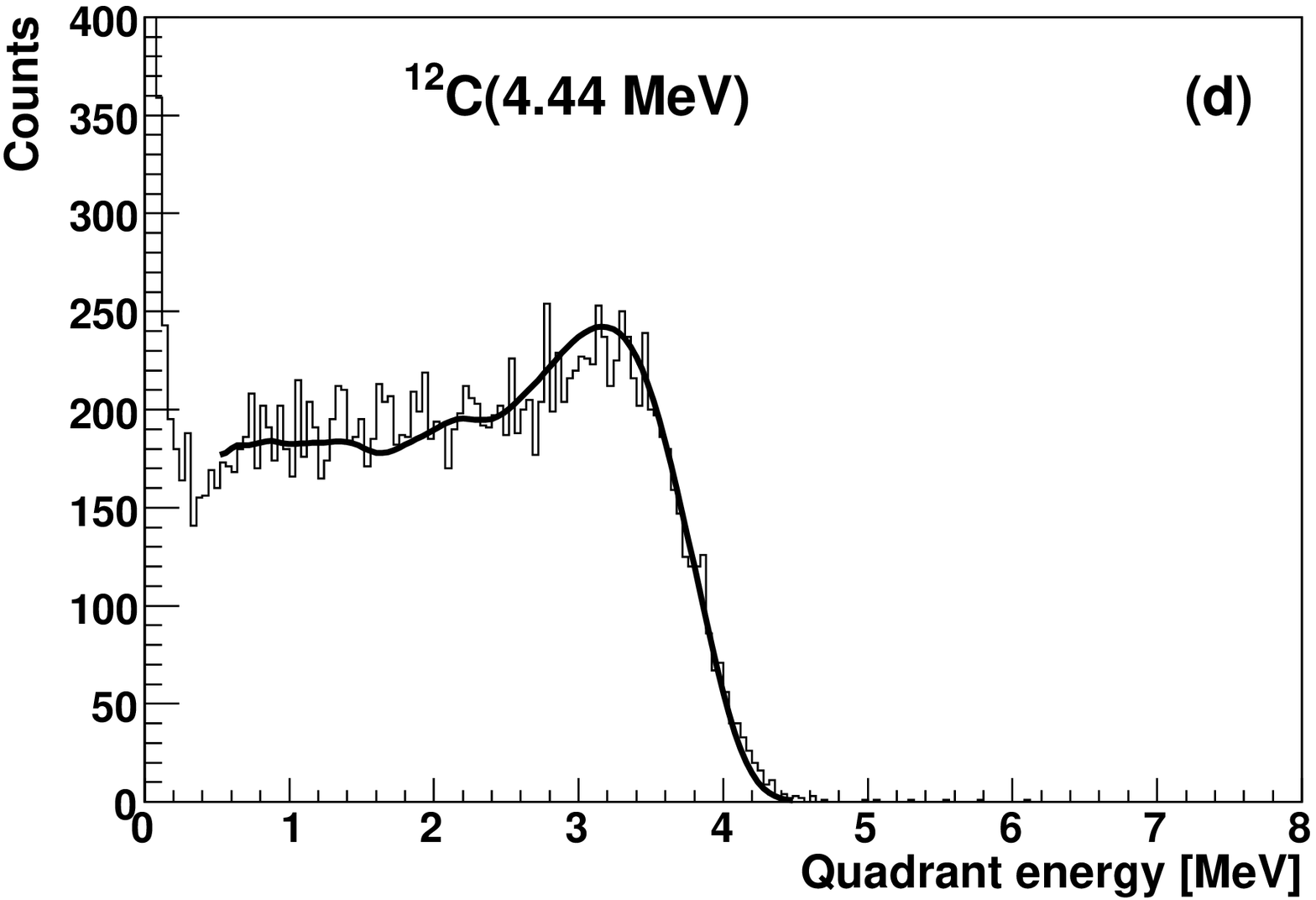}}
}
\centerline{
\mbox{\includegraphics[width=0.497\textwidth]{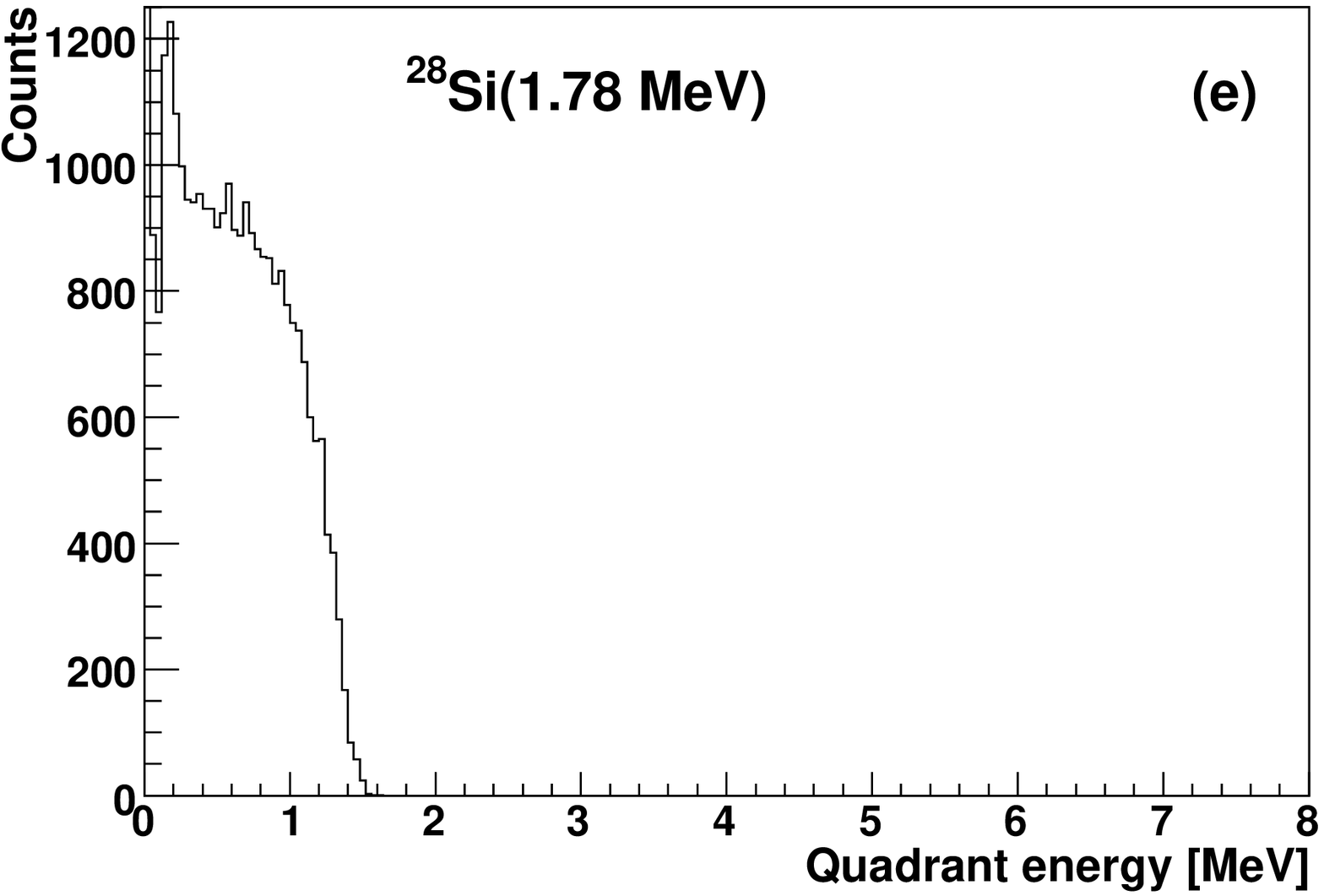}}
\mbox{\includegraphics[width=0.497\textwidth]{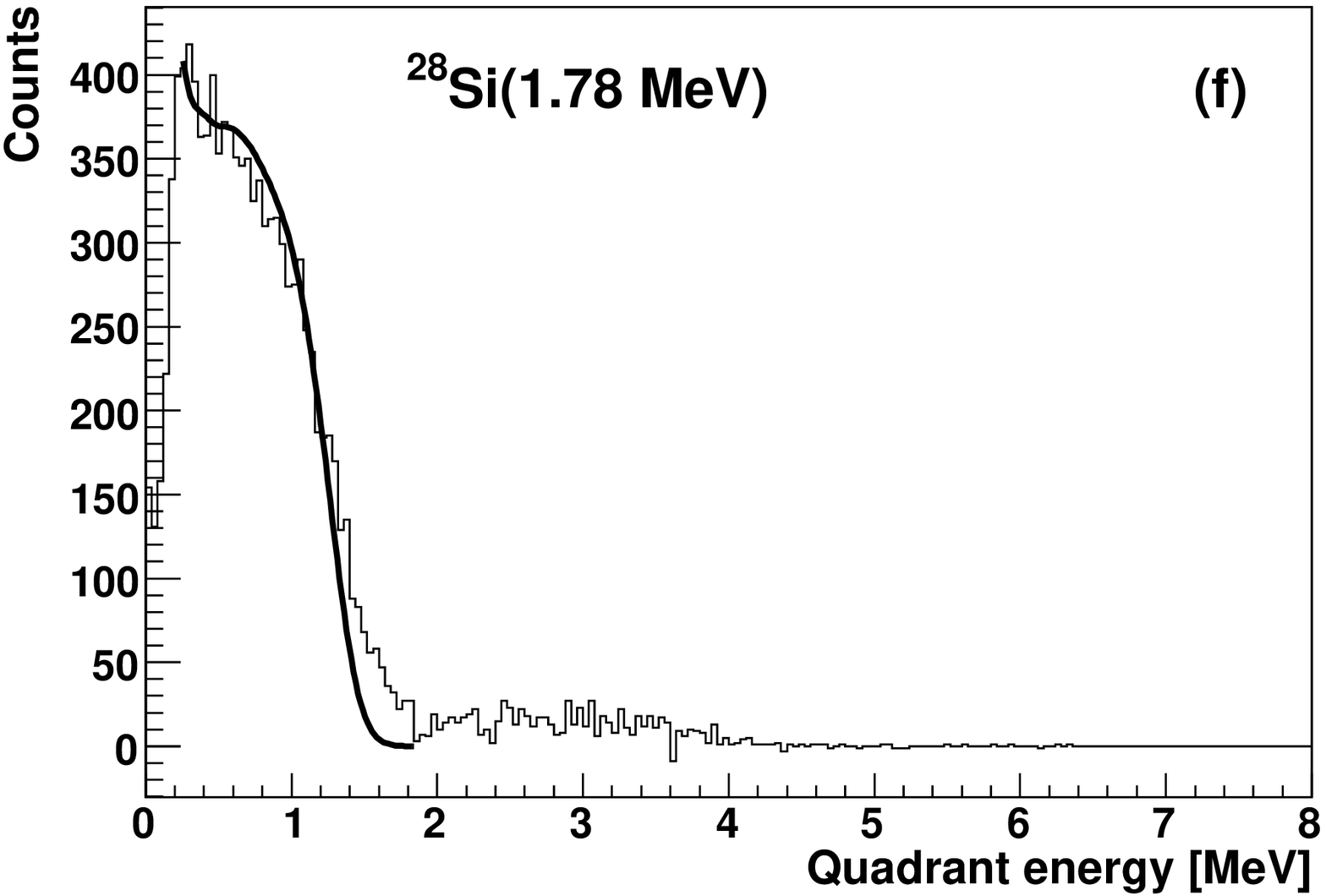}}
}
\caption{Left: unsmeared simulated 1-quadrant gamma-ray spectra. Right: the   
corresponding experimental spectra for gamma rays of energy 
(a),(b):1.17+1.33 MeV ($^{60}$Co), (c),(d):4.44 MeV ($^{12}$C), 
(e),(f):1.78 MeV ($^{28}$Si). A threshold of 0.16 MeV for 
the tube and of 0.13 MeV in at least one of the quadrants has been applied to
the simulated spectra. The thick dark lines show a fit to the
data using an empirically determined detector resolution function for 
the quadrants (see text).}
\label{comparison}
\end{figure}

The design goals for the detector were that it be able to stop and
measure the energies of the positrons and electrons produced from the
$^{12}$C(0$^+_2$) pair decay, while remaining relatively insensitive to the
gamma-ray cascade decay of that state.  Figure~\ref{stoppingRange}
illustrates the position within the detector,
where positrons or electrons from E0 transitions from the (a)
$^{12}$C(0$^+_2$; 7.65 MeV) and (b) $^{16}$O($0^+_2$; 6.05 MeV) states
stop.  Each figure shows 
the results of 10$^4$ e$^+$-e$^-$ pairs emitted at the target position in the 
center of the plot.  Electrons and positrons stopping in the two central 
tubes as well as in the scintillator cube are clearly identified. On both 
figures the effects of the Ta liners for the beam entrance and exit 
can be seen (between x=$\pm$5 mm). 

Most of the electrons and positrons from the 7.65 MeV $0^+_2$ state 
in $^{12}$C which enter the quadrants, stop near the center of the 
detector and away from the corners where light collection presumably 
deteriorates. Those positrons or electrons that exit the detector without 
depositing their full energy mainly do so through the central hole, the 
beam entrance and exit holes, the empty spaces between the 
acrylic and the scintillator tubes or the scintillator tube and 
the quadrants; some particles exit the detector after depositing energy 
in a quadrant.

Figures ~\ref{Edep}(a) and ~\ref{Edep}(b) show the energy deposited in the
active scintillator tube and the total energy deposited in all active
components of the detector (i.e., the scintillator tube and the four
quadrants) for simulated $^{16}$O(0$^+_2$) pair-decay events.  
The low-energy peak structure observed in Fig. ~\ref{Edep}(a) arises from
events where one of the two particles never reaches the tube, either
because it exited the device or was absorbed by the inert central
tube.  The higher energy peak corresponds to events where both
particles deposit energy in the scintillator tube.  In
Figure~\ref{Edep}(b), the spectrum peaks at a total deposited energy of
about 4.18 MeV, showing that on average approximately 1 MeV of pair
kinetic energy is lost by particles traversing the inert central
tube.

To determine the simulated detection efficiency for e$^+$-e$^-$ pairs, 
we make the following assumptions: for an event in which the energy 
deposited in the scintillator tube was greater than 0.16 MeV, and the energy
deposited in at least one of the four scintillator quadrants was greater
than 0.13 MeV, corresponding to the respective experimental energy 
thresholds, the pair was detected.  Otherwise, the pair was
not detected. The ratio between the number of detected pairs and 
total number of emitted pairs gives the simulated detection efficiency, and 
was $\sim$85\% for the decay of the $^{16}$O(0$^+_2$) state, and  
$\sim$90\% for the  $^{12}$C(0$^+_2$) state. Since this difference is 
small, observation of pairs from $^{16}$O(0$^+_2$) provides 
a good baseline for an estimate of the experimental
efficiency for pairs from the $^{12}$C(0$^+_2$) excitation, which
cannot be determined directly.  While the experimental detection
efficiency depends on other factors in addition to the total deposited
energy, such as timing, detector thresholds, etc., the 
difference between the efficiencies for the two pair transitions
should be relatively insensitive to these detailed effects.

\begin{figure*}
\centerline{
\mbox{\includegraphics[width=0.497\textwidth]{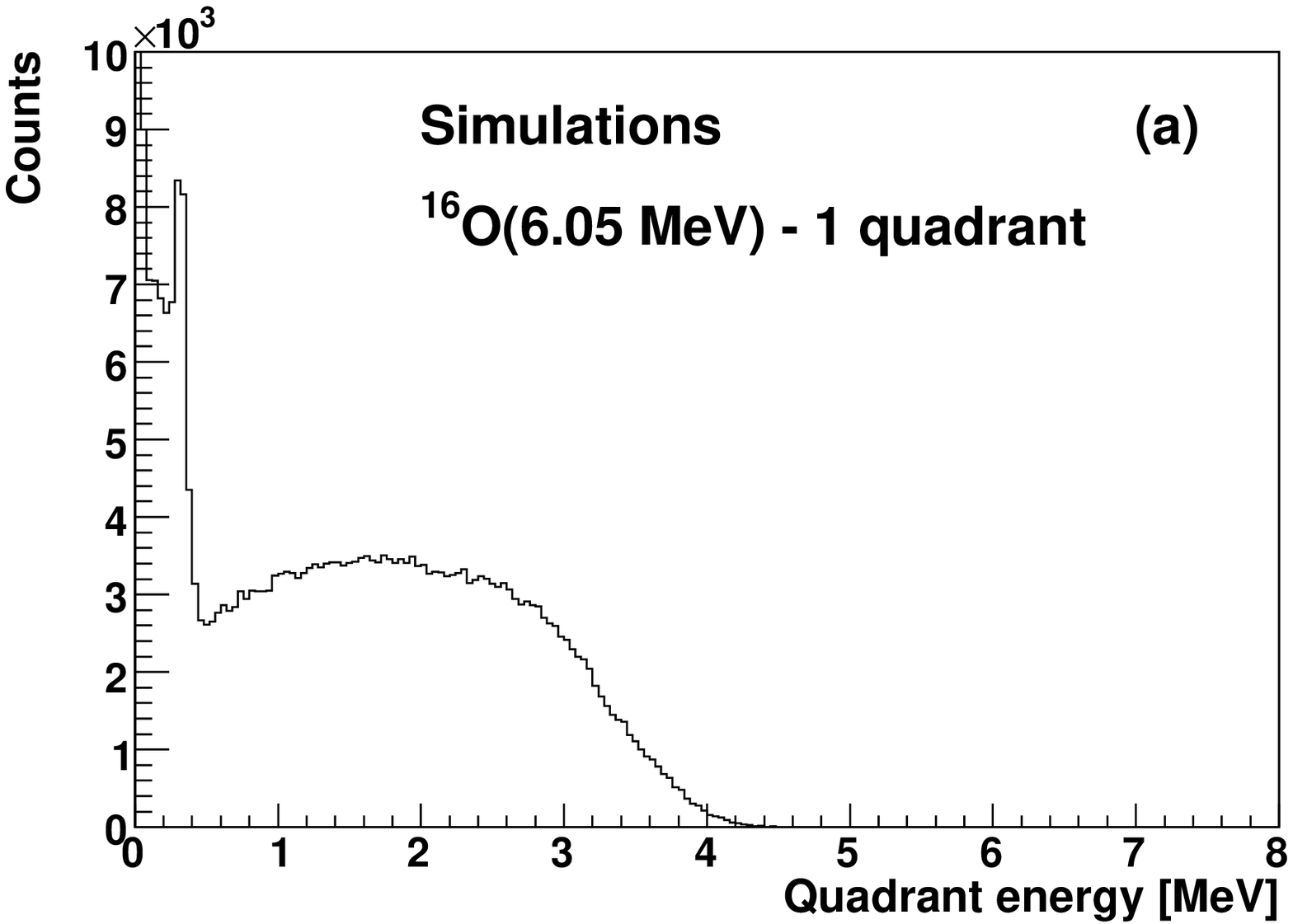}}
\mbox{\includegraphics[width=0.497\textwidth]{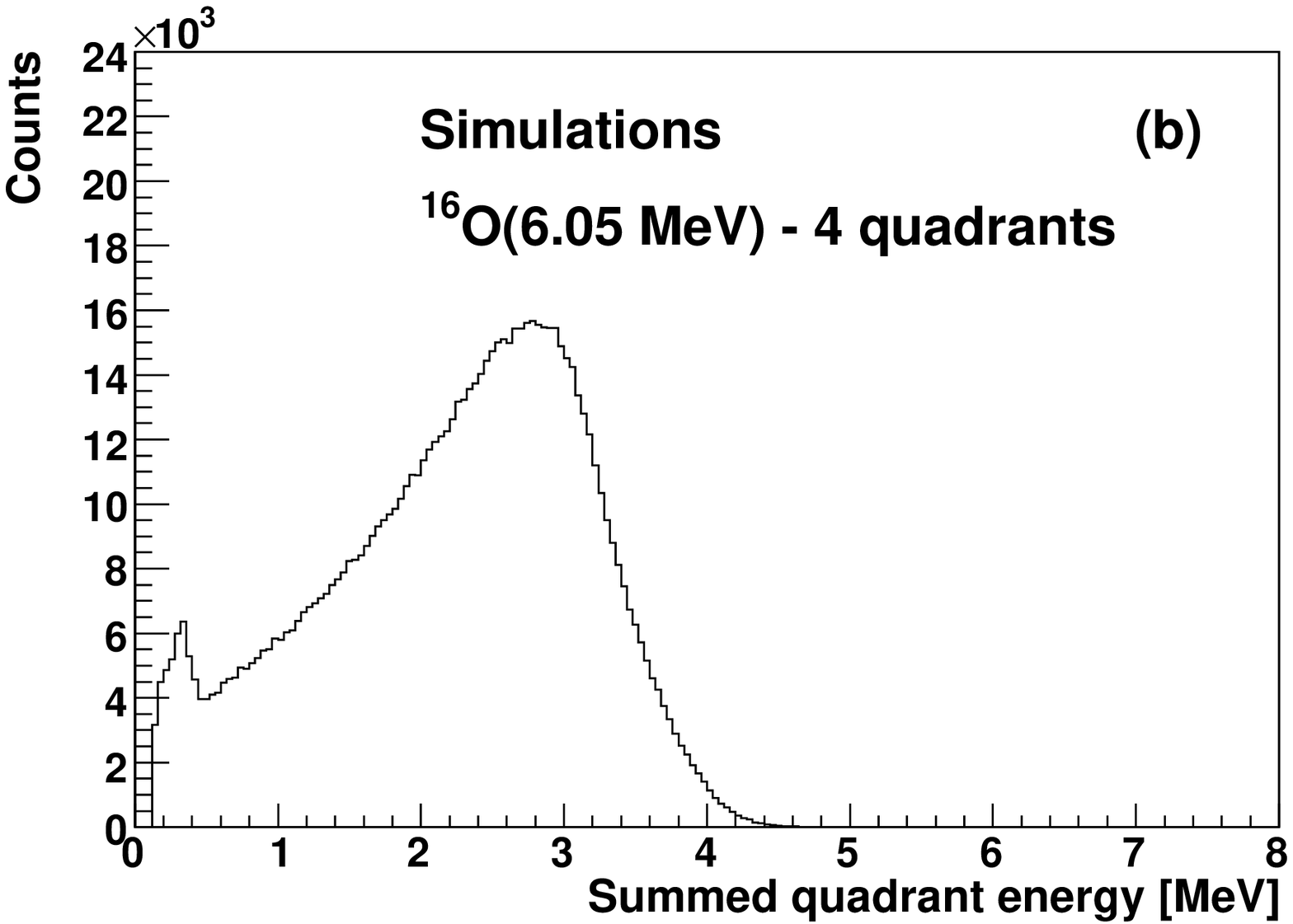}}
}
\centerline{
\mbox{\includegraphics[width=0.497\textwidth]{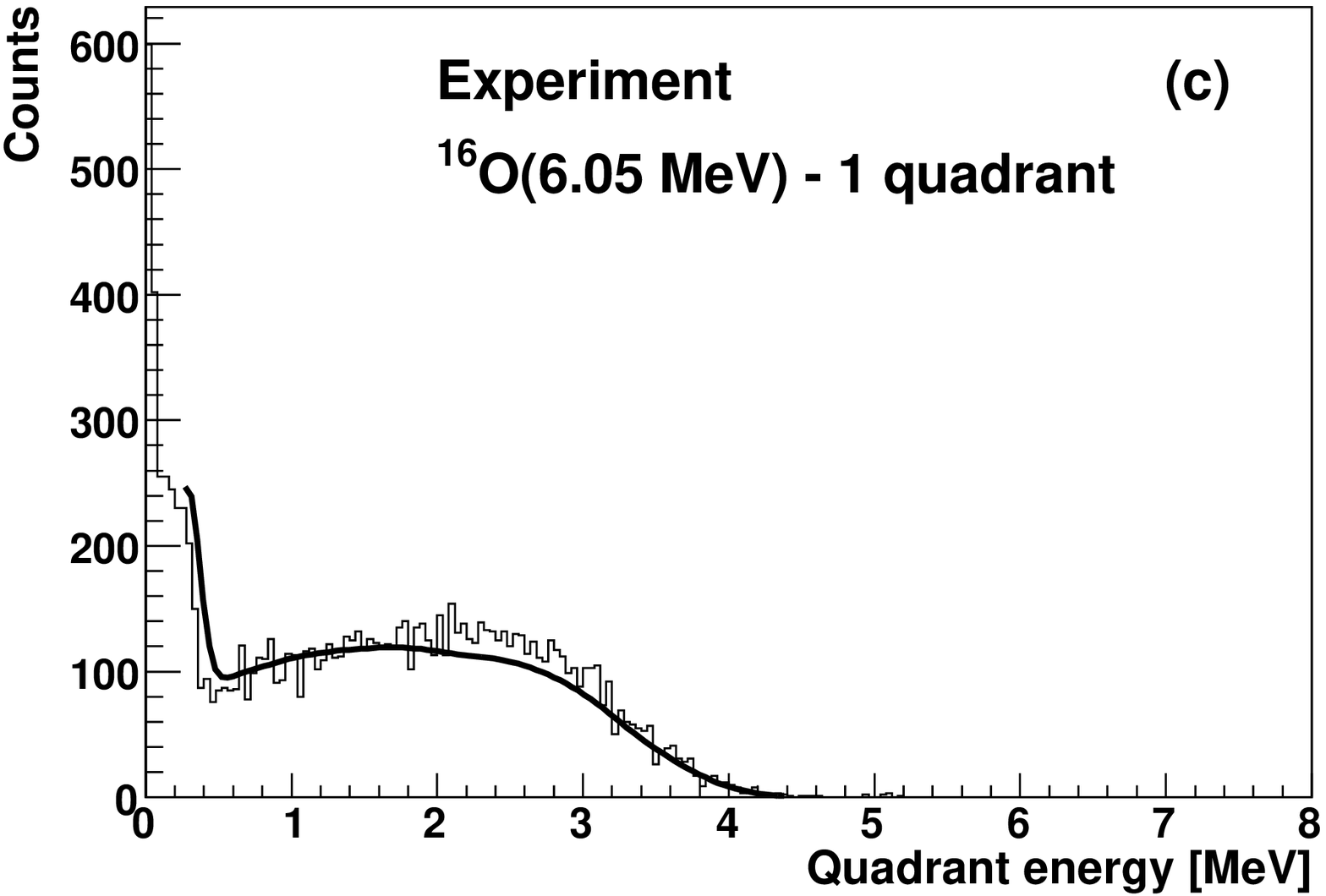}}
\mbox{\includegraphics[width=0.497\textwidth]{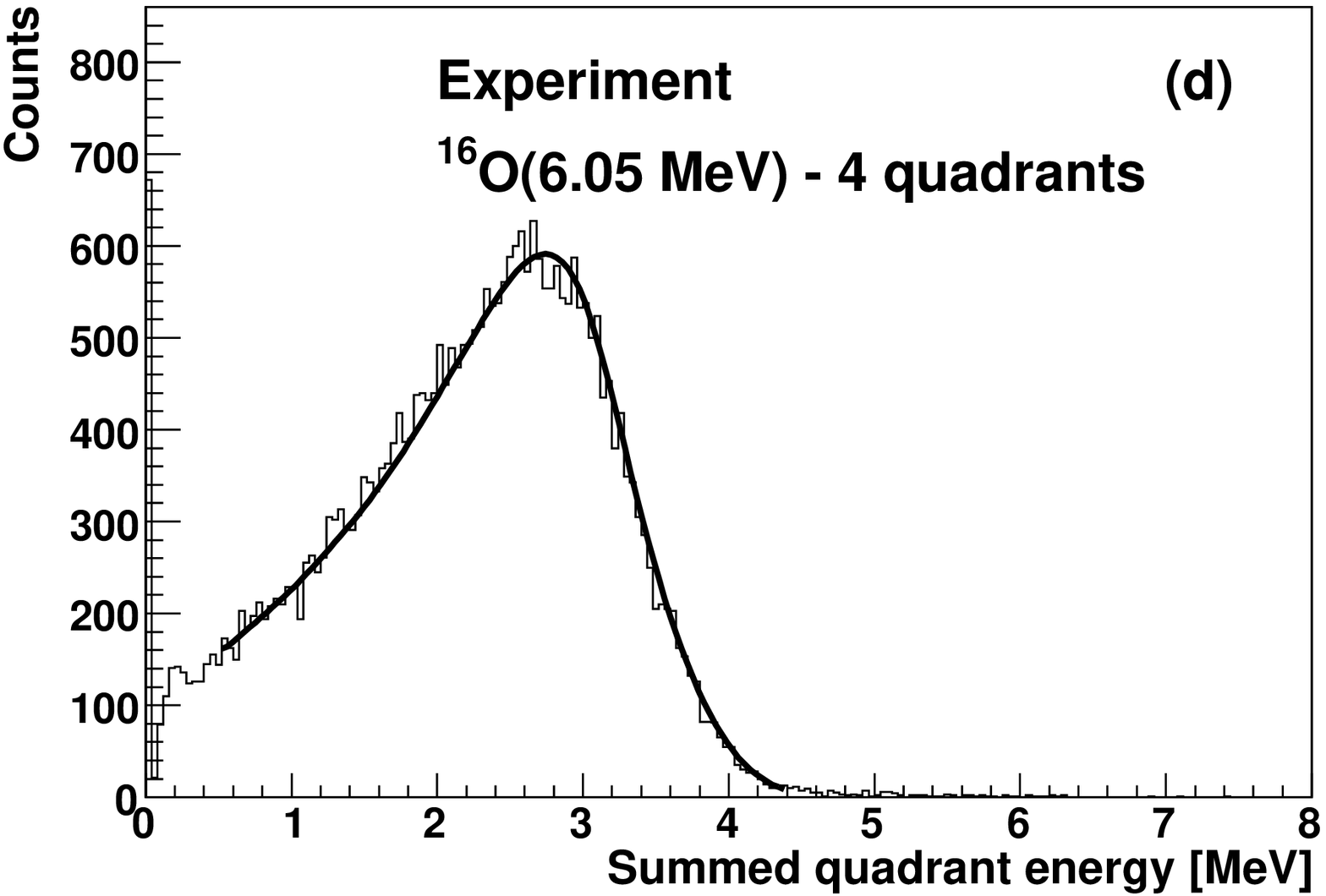}}
}
\caption{Spectra for (a),(b) the unsmeared simulations
and (c),(d) the experiment for energy deposited by electron-positron pairs 
from the $^{16}$O(0$^+_2$) in 1 quadrant ((a),(c)) and 
in all 4 quadrants ((b),(d)). In (c) and (d), the thick dark line shows a 
fit to the experimental spectra of the corresponding simulated spectra 
smeared with the appropriate detector resolution function (see text).}
\label{oxygenFits}
\end{figure*}

\subsection{Response to gamma rays}

\begin{figure*}
\centerline{
\mbox{\includegraphics[width=0.8\textwidth]{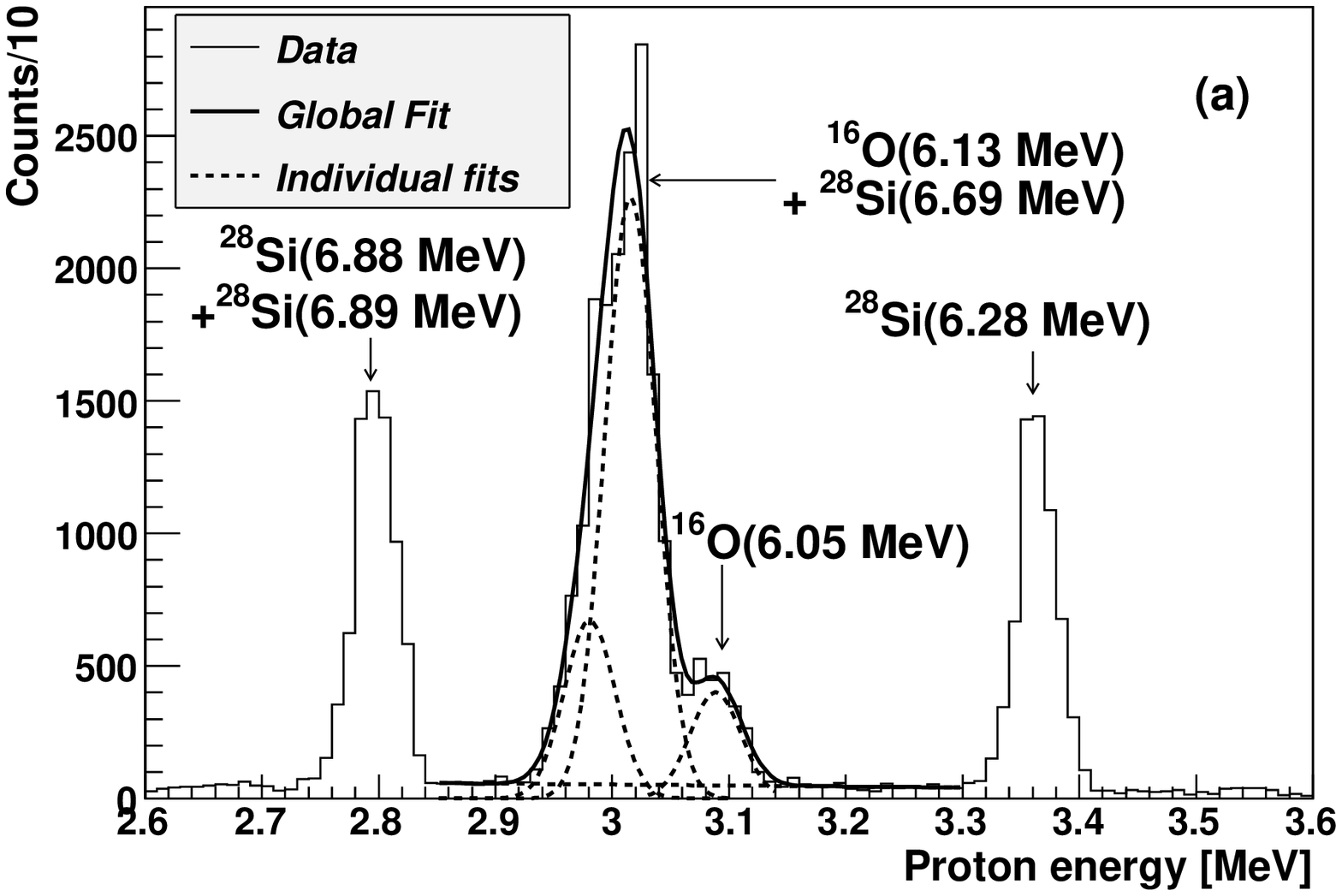}}
}
\centerline{
\mbox{\includegraphics[width=0.8\textwidth]{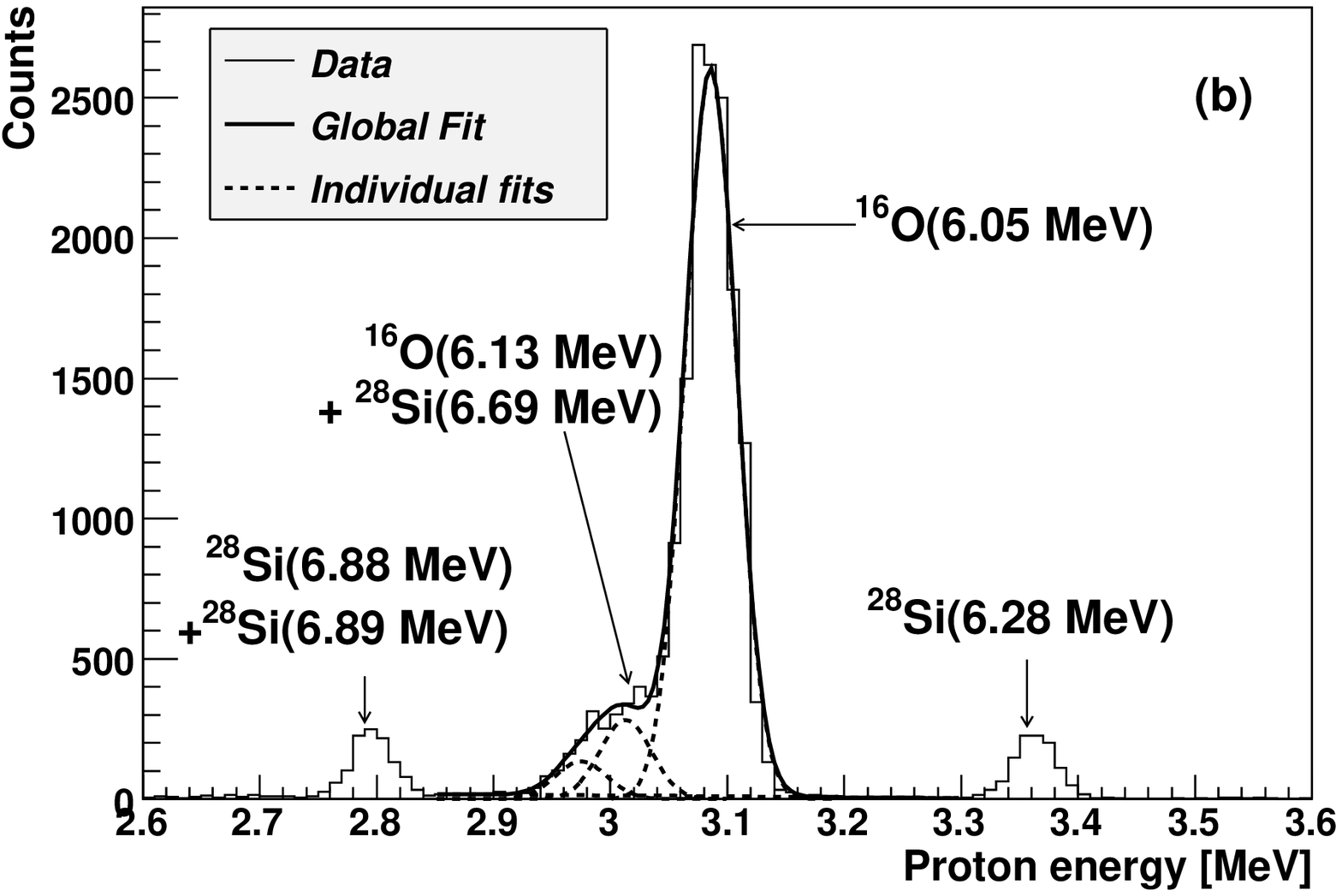}}
}
\caption{Proton-energy spectra from one segment of one of the silicon 
detectors for (a) singles and (b) coincidence events for the SiO$_{2}$ target. 
The singles spectrum has a down-scale factor of 10. The individual components
of the fit include a linear background and three Gaussian peaks corresponding 
to $^{28}$Si(6.69 MeV) at E$_p$=2.99 MeV,  $^{16}$O(6.13 MeV) at 
E$_p$=3.015 MeV, and $^{16}$O(6.05 MeV) at E$_p$=3.086 MeV.}
\label{ESi12}
\end{figure*}

Figure~\ref{gammaSims} shows the simulated total energy deposited
in the scintillator tube and all active detector elements for several 
gamma-ray transitions as detailed in the caption.  
The sensitivity of the device to the 511 keV photons is very 
small and therefore an additional 1.022 MeV of the single gamma-ray
transition energy is lost for many pair production events. The sharp peaks  
1.022 MeV below the maximum deposited energy in the total energy spectra 
illustrate this effect. 

The efficiency for gamma rays estimated from these simulations was
obtained in the same fashion as described above for e$^+$-e$^-$ pairs.  The
simulated efficiencies for single gamma-ray and cascade gamma-ray transitions 
are shown in Figure~\ref{gammaEff}, where they are compared to their 
experimental values. The experimental gamma-ray coincidence efficiency 
was determined by comparing proton events obtained with, and without 
coincident signals in the scintillator-detector array. For a given total 
gamma-ray energy, the efficiency for detecting at least one of the two 
photons in a cascade is systematically larger than that for a single 
photon carrying the same energy. Figure ~\ref{gammaSi} shows (a) a
"singles" proton-energy spectrum from p+SiO$_2$ collisions obtained
without the requirement of a scintillator coincidence, and (b) the
corresponding spectrum obtained with silicon-detector signals in
coincidence with pulses in the scintillator detectors, corrected for
random coincidences.  Several proton-energy peaks are observed
corresponding to different excitations in $^{16}$O and $^{28}$Si, as
well as in $^{12}$C that was present as an impurity in the target.
Clearly visible in Figure ~\ref{gammaSi}(b) is the relative enhancement
of the peak of the pair-emitting $^{16}$O($0^+_2$) excitation at
E$_p=3.086$ MeV compared to excitations that decay by gamma-ray emission.  
The experimental efficiency for e$^+$-e$^-$ pairs is discussed below.

\section{Comparison Between Experimental Results and Simulations}

\subsection{Energy calibrations}

Figure~\ref{calibration} illustrates
the method used to calibrate the tube energy. The peaks in the tube spectra
near 0.5 MeV ($^{12}$C(4.44 MeV)) and 1.2 MeV ($^{16}$O(6.05 MeV))
are first assigned a reasonable channel number
based on the corresponding uncalibrated experimental spectra. The 
calibration constants are then refined recursively by fitting the 
simulated spectra to data. Detector thresholds and an approximate detector 
resolution function are applied to the simulations for each detector 
component based on experimental data. Once the energy calibration
has been determined, the best fit parameters for the resolution function used 
to smear the simulated data are obtained by fitting the tube
spectrum for $^{16}$O(6.05 MeV). They are then fixed to fit the tube 
spectra for many gamma rays (we illustrate the case of $^{12}$C(4.44 MeV) 
on Figure ~\ref{calibration}(a)). The same method is used to determine the 
best fit parameters for the quadrants. This gives the following detector
resolution functions: a Gaussian with an energy dependent width of 
$\sigma (MeV) = 0.00097 + 0.16\times\sqrt{E} + 0.0010\times E$ for the tube 
spectra and $\sigma (MeV) = 0.0045 + 0.088\times\sqrt{E} + 0.00098\times E$ 
for the quadrant spectra, where E is the deposited energy. The widths of the 
resolution functions include three terms: an energy-independent part to 
account for the baseline noise, a $\sqrt{E}$ term to account for the 
statistical nature of light collection, and a term proportional to E to 
account for the variation of light collection efficiency within 
the detector. Figures~\ref{comparison}(b) and ~\ref{comparison}(d) show 
the same two-point calibration technique applied to one quadrant using 
the $^{12}$C(4.44 MeV) and the $^{60}$Co(1.17+1.33 MeV) transitions. 
The high-energy drop-off on the quadrant spectra occurs below the Compton 
edge for gamma rays from $^{12}$C and $^{28}$Si due to the requirement 
of energy being also deposited in the scintillator tube.  The simulations 
for $^{28}$Si(1.78 MeV) are convoluted using the same resolution function 
and normalized to data as shown on Figure~\ref{comparison}(f). The results
are in good agreement with the data.

Figure~\ref{comparison} also shows a comparison between the simulated 
(unsmeared) and the random-subtracted data spectra for various gamma-ray 
transitions after energy calibration has been performed. Figures 
\ref{oxygenFits}(a) and  \ref{oxygenFits}(b) show the unsmeared simulated 
spectra for energy deposited by e$^+$-e$^-$ pairs for the decay of 
$^{16}$O(6.05 MeV) in one quadrant and all four quadrants, respectively.
Resolution smearing is not included in Figures \ref{comparison}(a), 
\ref{comparison}(c), \ref{comparison}(e), and Figures 
\ref{oxygenFits}(a) and \ref{oxygenFits}(b); resolution smearing is, however, 
performed before normalization to data.
Figures \ref{oxygenFits}(c) and  \ref{oxygenFits}(d) show the corresponding
experimental energy spectra normalized to the simulations. The same detector 
resolution function as the one used for the gamma-ray quadrant spectra
has been applied to the simulated spectra before 
normalization; the results are in good agreement with the data.

\subsection{e$^+$-e$^-$ detection efficiency and gamma-ray suppression}

The experimental efficiency of the scintillator-detector array for
e$^+$-e$^-$ pairs from the $^{16}$O($0^+_2$) level was obtained in a
manner similar to that described above for gamma rays.
Fig.~\ref{ESi12} shows proton-energy spectra in the region of the
$^{16}$O($0^+_2$) peak for (a) singles proton events and (b)
random-subtracted coincidence events.  A complication in the
measurement arises from the nearby 3$^-$ excitation in $^{16}$O at 
E$_p$= 3.015 MeV and a 0$^+$ excitation in $^{28}$Si at E$_p$= 2.99 MeV,
which are not completely resolved from the $^{16}$O($0^+_2$) peak.  To
extract the yield for the $^{16}$O($0^+_2$), the data were fit with
three overlapping Gaussians.  The curves in Fig.~\ref{ESi12} show the
resulting fits to the proton-energy data.  In Fig.~\ref{ESi12}(a) the 
sum of the 3$^-$ ($^{16}$O) and 0$^+$ ($^{28}$Si)
peaks is dominant. With the application of the
scintillator-detector coincidence the $^{16}$O($0^+_2$) peak is
strongly enhanced relative to those corresponding to excitations that
decay by gamma rays.  The experimental value of the pair efficiency is
65.4 $\pm$ 1.8~\%, compared to the simulated value of 85\%.  The systematic 
over-prediction of the detector efficiency is similar to that observed for 
various gamma-ray transitions.The difference between the observed and 
simulated efficiencies is attributed to effects not treated in the 
simulation, such as light collection in the scintillators and light 
guides and a realistic treatment of electronic thresholds.  While 
these effects are not fully addressed in the simulation, the experimental 
efficiency may be calibrated by the $^{16}$O($0^+_2$) pair measurement, and
extrapolated to the nearby $^{12}$C($0^+_2$) state with guidance from
the Monte Carlo calculation.

\section{Conclusion}

The pair detector described here was used for an experiment to better
determine the rate of the triple alpha experiment by improving the error bars 
on the internal pair emission branch of the 0$^+_2$ excited state in $^{12}$C. 
The detector performance was simulated using the GEANT4 simulation framework. 
The simulations helped determine the best detector geometry 
at the design stage and were crucial for the detector 
calibration and the determination of important detector efficiencies. 
Monte Carlo simulations combined with empirically determined detector 
resolution functions yielded good agreement with the experimental data.

\section{Acknowledgements}
We wish to acknowledge the skilled assistance of Len Morris and Doug
Miller in the mechanical design and construction of the detector.  This
research was supported in part by the US National Science Foundation
Grants PHY01-10253 and PHY02-16783, the latter funding the Joint
Institute for Nuclear Astrophysics (JINA) and by US Department of
Energy Grant DE-FG02-04ER41320 (WMU).

\end{document}